\documentclass[9pt,conference]{IEEEtran}
\IEEEoverridecommandlockouts 
\usepackage{blkarray}                                      
\usepackage{algpseudocode}                                 
\usepackage{algorithm}

\usepackage{graphicx}                                      
\usepackage{amsmath}
\usepackage{amssymb}
\usepackage{amsfonts}
\usepackage{amsthm}
\usepackage[mathcal]{eucal}
\usepackage{mathrsfs}
\usepackage{booktabs}
\usepackage{enumerate}
\usepackage{multirow}
\usepackage[subrefformat=parens,farskip=0pt,justification=centering]{subfig}
\usepackage{color}
\usepackage{cite}                                          
\usepackage{comment}                                       
\usepackage{soul}                                          
\soulregister\cite7
\soulregister\ref7
\soulregister\pageref7
\usepackage{etoolbox}                                      
\usepackage{url}
\usepackage{nth}                                           
\usepackage{bm}                                            
\usepackage{courier}
\usepackage{makecell}
\usepackage{balance}
\usepackage{threeparttable}
\usepackage[bookmarks=false]{hyperref}
\hypersetup{
    colorlinks = true,
    citecolor  = blue,
    linkcolor  = blue,
    urlcolor   = blue,
}
\usepackage{tikz}
\usetikzlibrary{positioning,calc,fit,decorations.pathmorphing,shapes.geometric, shapes.gates.logic.US, calc}
\usetikzlibrary{arrows,arrows.meta,decorations.markings,shapes,shapes.arrows}
\usetikzlibrary{backgrounds,patterns}
\usepackage{pgfplots}
\usepackage{pgfplotstable}
\usepackage{scalefnt}
\usepgfplotslibrary{groupplots}
\pgfplotsset{compat=newest}
\usepackage{caption}
\usepackage{cleveref}
\Crefformat{figure}{Fig.~#2#1#3}                           
\Crefname{subfigure}{Fig.}{Figs.}
\Crefname{figure}{Fig.}{Figs.}
\Crefformat{table}{TABLE~#2#1#3}                           
\definecolor{CUpurple}{RGB}{117,15,109}
\definecolor{CUgold}{RGB}{221,163,0}
\definecolor{CUribbon}{RGB}{244,223,176}
\definecolor{CUblack}{RGB}{34,24,21}
\definecolor{USTgold}{RGB}{153,102,0}
\definecolor{USTyellow}{RGB}{204,153,0}
\definecolor{USTyellowlight}{RGB}{255,212,0}
\definecolor{USTorange}{RGB}{255,166,26}
\definecolor{USTblue}{RGB}{0,51,102}
\definecolor{USTmiddle}{RGB}{0,116,188}
\definecolor{USTlight}{RGB}{99,202,225}
\definecolor{USTgray}{RGB}{204,204,204}
\definecolor{USTred}{RGB}{237,27,47}
\definecolor{USTdarkred}{RGB}{124,35,72}

\usepackage{tcolorbox}
\tcbuselibrary{skins,breakable}
    {\endtcolorbox}

\usepackage{titlesec}

\crefname{mytheorem}{Theorem}{Theorems}
\crefname{mylemma}{Lemma}{Lemmas}
\crefname{myclaim}{Claim}{Claims}
\crefname{myproperty}{Property}{Properties}
\crefname{mycorollary}{Corollary}{Corollaries}

\algrenewcommand\textproc{\texttt}

\makeatletter
\let\OldStatex\Statex
\renewcommand{\Statex}[1][3]{%
  \setlength\@tempdima{\algorithmicindent}%
  \OldStatex\hskip\dimexpr#1\@tempdima\relax
}
\makeatother

\RequirePackage[normalem]{ulem} 
\RequirePackage{color}\definecolor{RED}{rgb}{1,0,0}\definecolor{BLUE}{rgb}{0,0,1} 

\usepackage{times}
\usepackage{upgreek}
\usepackage{breqn}
\usepackage{comment}
\usepackage{multirow}
\usepackage{graphicx}
\graphicspath{{./fig/}}

\newif\ifanonymousreview
\anonymousreviewtrue
\ifanonymousreview
    \hypersetup{pdfauthor={}}
\fi

\DeclareMathAlphabet\mathbfcal{OMS}{cmsy}{b}{n}

\usepackage{amsmath}

\definecolor{plotBlue}{RGB}{44,111,155}
\definecolor{plotOrange}{RGB}{189,108,38}
\definecolor{plotRed}{RGB}{184,91,81}
\definecolor{plotPurple}{RGB}{120,86,154}
\definecolor{plotTeal}{RGB}{77,140,124}

\definecolor{myblue}{RGB}{73,148,196}    
\definecolor{mydarkblue}{RGB}{18,38,79} 
\definecolor{myorange}{RGB}{234,85,20}  
\definecolor{myyellow}{RGB}{250,192,61}   
\definecolor{mypink}{RGB}{252,228,215}   

\begin{document}

\title{
LLM-Assisted Co-Optimization of RTL and Synthesis Algorithms via Bottleneck Analysis
}

\author{
    Yikang Ouyang$^1$, \quad
    Yang Luo$^1$, \quad
    Dongsheng Zuo$^2$, \quad
    Xiaoxiao Liang$^1$, \quad
    Yuzhe Ma$^{1\dagger}$ \\
    $^1$The Hong Kong University of Science and Technology (Guangzhou) \\
    $^2$The Chinese University of Hong Kong
    \thanks{$^{\dagger}$Corresponding author}
}

\maketitle
\begin{abstract}

Post-synthesis power, performance, and area (PPA) depend jointly on the register-transfer-level (RTL) code and the synthesis algorithms.
However, the benefits of RTL optimization may be limited by bottlenecks in synthesis algorithms, and vice versa, leaving opportunities for better PPA unexplored by methods that optimize either level alone.
In this work, we present a large language model (LLM)-assisted framework for regional co-optimization of RTL and synthesis algorithms via bottleneck analysis.
A key insight behind our approach is that RTL structures and the Boolean networks used in synthesis expose complementary opportunities for addressing bottlenecks that lead to suboptimal PPA in different regions of the circuit.
Our bottleneck analysis uses a Gaussian process (GP) to assess whether optimizing RTL or synthesis algorithms is more promising for each region.
The GP learns from regional graph representations and PPA feedback. 
Using the GP's predictions and predictive uncertainty, Bayesian optimization (BO) provides guidance that an LLM Router combines with circuit analysis to choose how to address each region's bottlenecks.
Self-improving LLM workers then optimize the RTL or synthesis algorithms in their assigned regions, and their regional changes are greedily integrated into the current design configuration.
Experiments on 16 RTL designs show average improvements of 103.27\% in worst negative slack (WNS), 1.23\% in area, and 13.26\% in power relative to the initial designs.
These improvements exceed those of the strongest baseline for each metric by 47.80, 2.75, and 13.17 percentage points, respectively.

\end{abstract}

\section{Introduction}
\label{sec:introduction}


Post-synthesis power, performance, and area (PPA) depend jointly on the quality of register-transfer-level (RTL) code and the effectiveness of synthesis algorithms.
Functionally equivalent RTL descriptions can be implemented with different structures, while synthesis algorithms influence how these structures are transformed into gate-level netlists.
Extensive research has therefore explored optimization at both levels. RTL optimization restructures arithmetic and control logic to obtain implementations with better post-synthesis PPA.
Rule-based approaches use PPA feedback to guide the search for functionally equivalent RTL implementations~\cite{RTL-DAC2023-Coward,RTL-TCAD2024-Coward}.
More recently, large language models (LLMs) have enabled automated RTL optimization guided by the semantics of circuit operations~\cite{RTL-ICCAD2024-Yao,RTL-NEURIPS2025-Wang}.
Notably, RTLScout~\cite{COOPT-ARXIV2026-Arnold} rewrites RTL and replaces selected RTL fragments with synthesized netlists.
However, it does not try to propose more powerful synthesis algorithms, thus we also classified it as an RTL optimization method.
At the synthesis level, algorithms optimize Boolean structures through transformations such as rewriting and resubstitution, followed by technology mapping~\cite{LS-DAC2006-Mishchenko,LS-ICCAD2007-Mishchenko}. 
Advances in window-based resynthesis further expand the exploration of alternative logic structures~\cite{LS-ASPDAC2022-Riener}.
Beyond developing individual algorithms, recent studies have investigated automated synthesis algorithm optimization with the help of LLMs~\cite{LS-DAC2026-Yu}.

\begin{figure}[!t]
    \centering
    \includegraphics[width=\columnwidth]{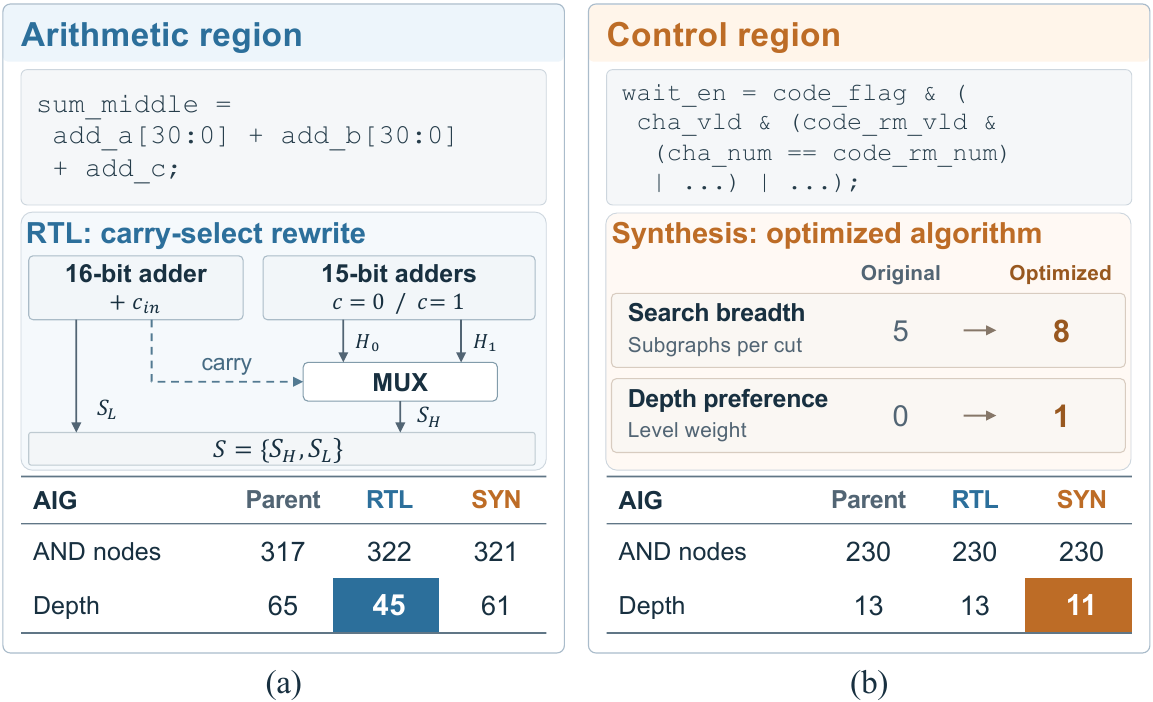}
   \caption{An example of RTL and synthesis algorithm optimization of two regions in an \texttt{arm\_cpu2} design: (a) arithmetic and (b) control regions respond differently to optimization at the two levels. 
   Tables report pre-mapping and-inverter graph (AIG) depth and AND-node counts for the two regions.}
    \label{fig:optimization-overview}
\end{figure}

Despite progress at both levels, the benefits of optimizing RTL or synthesis algorithms alone may be limited by unresolved bottlenecks at the other level. 
Here, a \textbf{bottleneck} is a structure in the RTL or Boolean network that leads to suboptimal timing, area, or power.
Such structures may arise from inefficient arithmetic or control organization in RTL, or persist because the complexity of the resulting Boolean networks makes them difficult for synthesis algorithms to optimize effectively.
Different parts of the same design may contain distinct bottlenecks and respond differently to optimization at the two levels. 
The arithmetic and control computations in~\Cref{fig:optimization-overview} illustrate this situation.

The example considers two \textbf{regions} of an \texttt{arm\_cpu2} design, each comprising an RTL code segment and its corresponding subgraphs in the control/data-flow graph (CDFG) and the pre-optimization Boolean network (in AIG format).
The formal definition of a region is given in~\Cref{sec:graph-representations}.
We refer to RTL rewriting and synthesis algorithm optimization as the RTL and SYN optimization branches, respectively.
In the arithmetic region computing \texttt{sum\_middle}, a carry-select RTL rewrite changes the arithmetic organization, reducing AIG depth from 65 to 45 levels. 
However, SYN optimization of the same region yields a smaller depth reduction to 61 levels, since RTL adder naturally forms deep carry-propagation logic if the RTL structure is not changed, which cannot be effectively optimized by synthesis algorithms like rewrite alone~\cite{ADDER-TCAD2019-Ma}.
In the control region computing \texttt{wait\_en}, the SYN worker edits ABC's rewriting heuristic, raising the candidate subgraph limit from five to eight and the level weight from zero to one. With these edits, balancing reduces AIG depth from 13 to 11 levels, while the RTL rewrite leaves it at 13.
These outcomes motivate us to examine, for each region, whether rewriting the RTL or modifying the synthesis algorithm would yield the larger whole-design PPA gain.

To examine such optimization potential at both RTL and SYN levels, we propose an LLM-assisted framework for co-optimization of RTL and synthesis algorithms via bottleneck analysis.
We represent each region using its CDFG and AIG to capture computational semantics and logic structure.
A Gaussian process (GP) learns from graph neural network (GNN) embeddings of these graphs and the whole-design PPA gains of past trials to predict the gain of each region--branch pair.
Bayesian optimization (BO) ranks pairs by predicted gain and uncertainty. An LLM Router combines these guidance with its analysis of the regional graphs' semantics to assign  each regions with a branch to perform optimization.
During optimization, RTL  LLM workers rewrite the selected RTL, while SYN LLM workers can reorder synthesis passes or edit the heuristics implemented within an ABC pass.
Both types of workers are self-improving LLM agents with a reviewer LLM that uses PPA feedback to refine their optimization knowledge for subsequent rounds, consolidating successful strategies and suppressing ineffective ones.
The resulting regional changes are greedily integrated into the current design configuration to improve whole-design post-synthesis PPA.

The main contributions of this work are as follows:
\begin{itemize}
    \item We propose an LLM-assisted framework that co-optimizes RTL and synthesis algorithms to improve post-synthesis PPA.

    \item We develop a bottleneck analysis method that combines LLM reasoning over regional RTL and Boolean structures with GP-guided Bayesian optimization.
     By incorporating structural information and optimization feedback, the method assigns promising optimization branches to regions with bottlenecks.

    \item We develop LLM-based RTL and SYN workers for RTL rewriting and synthesis algorithm optimization.
    Reviewer feedback refines the workers' optimization knowledge to improve their ability to optimize subsequent regions. Greedy integration accepts a regional change only if it reduces the current whole-design PPA score.

    \item Across 16 RTL designs, our framework achieves average improvements of 103.27\% in worst negative slack (WNS), 1.23\% in area, and 13.26\% in power relative to the initial configurations. These improvements exceed those of the strongest baseline for each metric by 47.80, 2.75, and 13.17 percentage points, respectively.
\end{itemize}

\section{Preliminaries}
\label{sec:problem}
We summarize frequently used notation in~\Cref{tab:notation}.

\subsection{RTL and Logic Synthesis}
\label{sec:optimization-domains}

RTL describes computation and control logic; functionally equivalent descriptions can expose different synthesis opportunities and yield different post-synthesis PPA~\cite{RTL-TCAD2024-Coward,RTL-ICCAD2024-Yao}.
RTL optimization rewrites arithmetic expressions and control logic while preserving functionality.
Rule-based methods search for equivalent implementations through predefined rewrites~\cite{RTL-DAC2023-Coward,RTL-TCAD2024-Coward}, while LLM-based methods such as RTLRewriter and Dr.\ RTL use synthesis feedback to guide iterative PPA optimization~\cite{RTL-ICCAD2024-Yao,RTL-ARXIV2026-Fang}.

Logic synthesis transforms RTL into a gate-level netlist through Boolean optimization and technology mapping~\cite{LS-CAV2010-Brayton}.
Traditional methods include AIG rewriting and balancing~\cite{LS-DAC2006-Mishchenko} and cut rewriting~\cite{LS-DATE2019-Riener}.
Recent work applies LLMs to EDA algorithm optimization~\cite{EDA-DAC2026-Ghose,PD-ARXIV2026-Zhu,PD-ARXIV2026-Liu}; in logic synthesis, Self-Evolved ABC uses LLMs and PPA feedback to improve synthesis algorithms~\cite{LS-DAC2026-Yu}.
These optimization levels are complementary: the RTL methods discussed above keep synthesis algorithms fixed, while the synthesis methods keep the input RTL fixed.



\subsection{CDFG and Boolean Network}
\label{sec:graph-representations}

\begin{figure}[!t]
    \centering
    \includegraphics[width=\columnwidth]{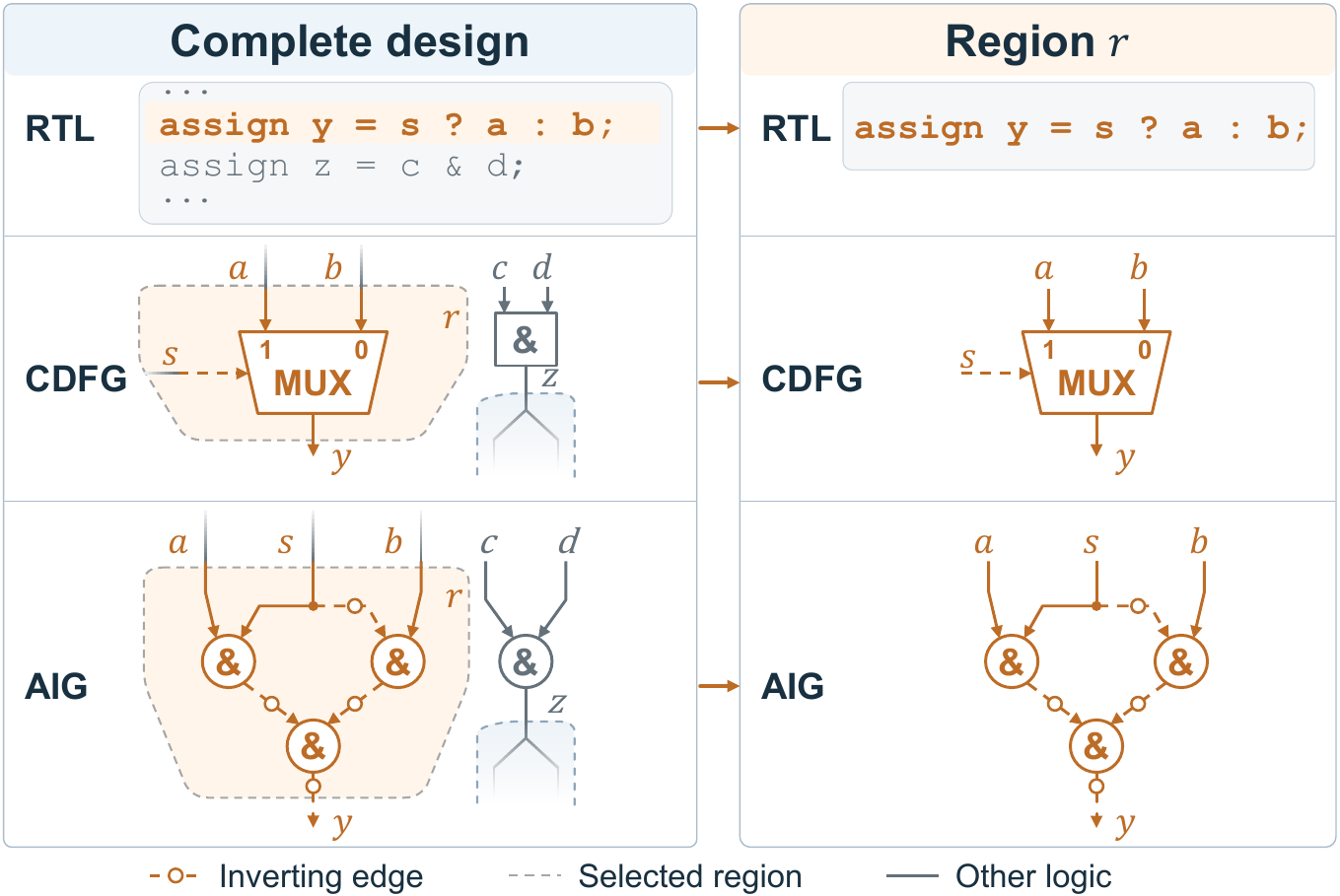}
    \caption{Example of a region represented by its RTL code and corresponding CDFG and AIG subgraphs.}
    \label{fig:region-representations}
\end{figure}

A \textbf{control/data-flow graph} (CDFG) captures RTL computations and their data and control dependencies.
CDFGs and related graph representations support dynamic behavior analysis~\cite{RTL-AAAI2026-Ma} and PPA estimation~\cite{RTL-ICML2026-Liu,RTL-ISCA2022-Xu,RTL-ICCAD2023-Fang}.
A \textbf{Boolean network} represents the logic derived from RTL during synthesis.
We use an \textbf{and-inverter graph} (AIG), which represents Boolean functions using two-input AND nodes and inverting edges.
AIG-based synthesis commonly uses algorithms such as rewriting and balancing~\cite{LS-DAC2006-Mishchenko}.
As illustrated in~\Cref{fig:region-representations}, a \textbf{region} consists of an RTL fragment and the subgraphs of the full design's CDFG and AIG that correspond to that fragment.

\subsection{Problem Formulation}
\label{sec:joint-formulation}

We jointly optimize RTL source $R_t$ and synthesis algorithms $A_t$ through the joint design configuration at round $t$:
\begin{equation}
    D_t=(R_t,A_t),\qquad
    A_t=[\alpha_1,\ldots,\alpha_{m_t}],
    \label{eq:design-state}
\end{equation}
where each $\alpha_i$ specifies a synthesis algorithm (e.g., \textit{rewrite}).
Synthesis produces the post-synthesis netlist $N_t$:
\begin{equation}
    N_t=\operatorname{Synthesis}(R_t,A_t).
    \label{eq:netlist-evaluation}
\end{equation}
We measure its worst negative slack (WNS) $W(N_t)$, total negative slack (TNS) $T(N_t)$, cell area $\mathrm{Area}(N_t)$, and power $P(N_t)$.
$\mathcal{R}_t$ denotes the candidate region set at round $t$.


\begin{table}[t]
\centering
\caption{ Frequently used notations.}
\label{tab:notation}
\small
\begin{tabular}{@{}p{0.30\columnwidth}p{0.65\columnwidth}@{}}
\toprule
Notation & Explanation \\
\midrule
$D_t=(R_t,A_t)$ & Design configuration at round $t$: RTL source $R_t$ and synthesis algorithms $A_t$ \\
$S(D)$ & PPA score used as the optimization objective \\
$N_t$ & Post-synthesis netlist of $D_t$ \\
$W,T,\mathrm{Area},P$ & Whole-design WNS, TNS, cell area, and power \\
$r\in\mathcal{R}_t$ & An RTL fragment and the corresponding subgraphs of the full design's CDFG and AIG. $\mathcal{R}_t$ is the set of candidate regions at round $t$ \\
$b\in\{\mathrm{RTL},\mathrm{SYN}\}$ & Optimization branch for region $r$ \\
$D'_{t,r,b}$ & Candidate configuration obtained by optimizing region $r$ through branch $b$ from $D_t$ \\
$g_t(r,b)$ & PPA score gain $S(D_t)-S(D'_{t,r,b})$ from regional optimization \\
\bottomrule
\end{tabular}
\end{table}

\begin{figure*}[!t]
    \centering
    \includegraphics[width=\textwidth]{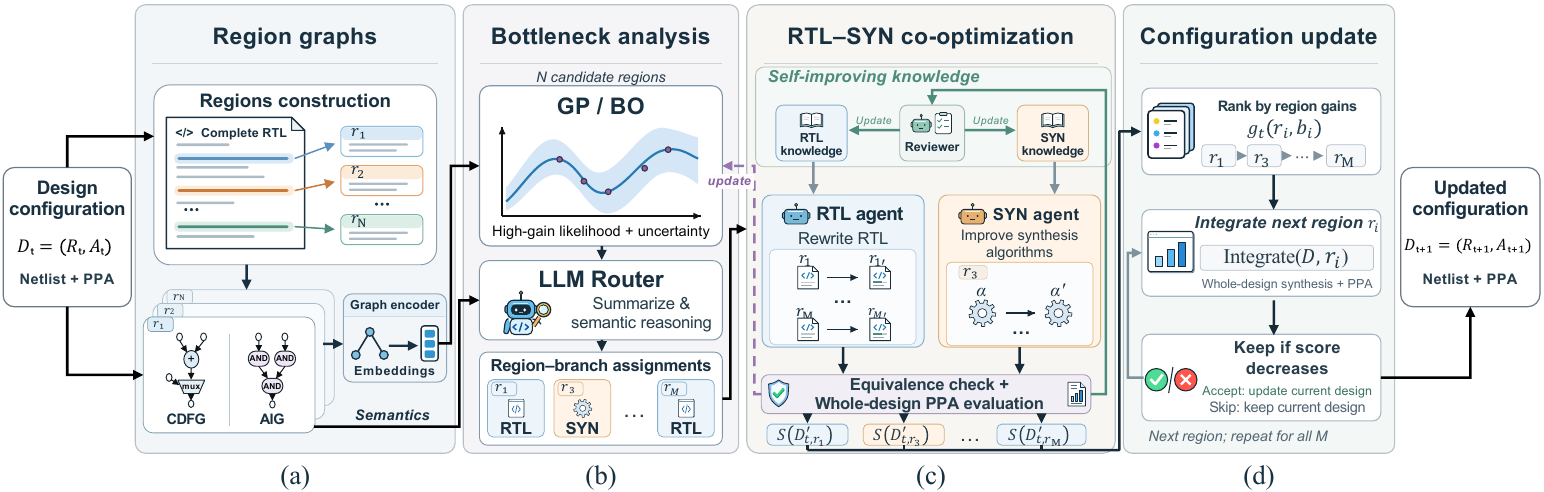}
\caption[RTL and synthesis algorithm co-optimization via bottleneck analysis]{LLM-assisted regional co-optimization of RTL and synthesis algorithms via bottleneck analysis.
(a) Candidate regions with associated CDFG and AIG subgraphs, which are encoded using separate GNNs.
(b) Active learning uses regional embeddings and PPA feedback to predict PPA gains with a GP and provide BO guidance based on predictions and uncertainty.
The LLM Router analyzes bottlenecks by combining this numerical guidance with semantic reasoning to assign regions to RTL/SYN branches.
(c) Self-improving worker LLMs optimize assigned regions from a common parent configuration. A reviewer uses the outcomes to refine their optimization knowledge.
(d) Regional optimizations are greedily integrated to obtain an updated design configuration.}
    \label{fig:method-overview}
\end{figure*}

Across rounds, we select a configuration minimizing the weighted PPA score:
\begin{equation}
\begin{aligned}
 D^* &\in \underset{D\in\{D_t\}_t}{\arg\min}\; S(D),\\
 S(D)&=\omega_W\frac{W-W_0}{W_0}
       +\omega_T\frac{T-T_0}{T_0}\\
       &\quad+\omega_A\frac{\mathrm{Area}-\mathrm{Area}_0}{\mathrm{Area}_0}
       +\omega_P\frac{P-P_0}{P_0},
\end{aligned}
\label{eq:score}
\end{equation}
where $W,T,\mathrm{Area},P$ are post-synthesis measurements for $D=(R,A)$, subscript $0$ denotes their values for the initial configuration $D_0$, and $\omega_W,\omega_T,\omega_A,\omega_P>0$ are preference weights.
We assume that the initial design violates timing, with $W_0<0$ and $T_0<0$.

\section{Bottleneck Analysis}
\label{sec:analysis}
Our framework identifies bottlenecks in design regions and assesses whether RTL rewriting or synthesis algorithm optimization is more promising for each region.
At round $t$, we perform bottleneck analysis using the post-synthesis netlist $N_t$, its associated design configuration $D_t=(R_t,A_t)$, and the PPA score $S(D_t)$ to identify bottlenecks in regions, as shown in~\Cref{fig:method-overview}~(a) and~(b).

\Cref{sec:representation} describes how we construct candidate regions and represent their computational and Boolean structures using CDFG and AIG subgraphs, respectively.
A Gaussian process (GP) learns from embeddings of these graphs and PPA feedback after regional optimization to assess the potential of each region--branch pair.
Bayesian optimization (BO) uses the GP's predictions and uncertainty to guide region--branch selection.
The LLM Router combines this guidance with structural and semantic reasoning over the regional graphs to make the final branch assignments (\Cref{sec:bottleneck}).



\subsection{Graph Representation for Bottleneck Analysis}
\label{sec:representation}

We model both the RTL computational structure and the AIG logic structure of each region to characterize optimization opportunities at the two levels.
We first convert the current RTL $R_t$ into a CDFG and partition it into logic-cone regions, forming the candidate region set $\mathcal{R}_t$.
Each region $r\in\mathcal{R}_t$ comprises a portion of RTL source code implementing that local logic cone, together with its corresponding CDFG subgraph $G_r^{\mathrm{CDFG}}$ and pre-optimization AIG subgraph $G_r^{\mathrm{AIG}}$.

We use separate graph encoders to model the CDFG and AIG structures.
The CDFG encoder captures RTL computation and control structure using operator type, bit width, signedness, and sequential roles as node attributes, with edges distinguishing data and control dependencies.
The AIG encoder captures Boolean logic structure through the connectivity of two-input AND nodes and edge inversion.
In each encoder, message passing aggregates information along graph edges, and mean pooling produces a regional embedding.
We train both encoders through self-supervision to reconstruct masked node attributes and predict graph statistics, including operator composition, depth, and fanout, without requiring PPA labels.

We concatenate the two embeddings into a unified structural representation:
\begin{equation}
    z_r=[h_r^{\mathrm{CDFG}};h_r^{\mathrm{AIG}}],
    \label{eq:region-representation}
\end{equation}
where $h_r^{\mathrm{CDFG}}$ and $h_r^{\mathrm{AIG}}$ encode the region's RTL and AIG structures, respectively.
The resulting representation $z_r$ provides complementary structural features to actively learn which optimization branches can reduce the PPA score, as later described in~\Cref{sec:bottleneck}.

\subsection{Bottleneck Analysis with GP/BO Guidance and LLM Reasoning}
\label{sec:bottleneck}

The GP uses regional graph embeddings and feedback from prior trials to predict the whole-design PPA score gain for each region--branch pair.
BO uses these predictions and their uncertainty to propose a branch for each region.
The LLM Router analyzes the region's RTL operations, data and control dependencies, and Boolean structure to assess how each branch could address its bottleneck.
Drawing on this semantic reasoning and the available BO guidance, the Router determines whether to assign the region to RTL rewriting or synthesis algorithm optimization.


Starting from the parent configuration $D_t$, optimizing region $r$ through branch $b$ produces a candidate configuration $D'_{t,r,b}$.
We synthesize this configuration and evaluate whole-design PPA to obtain the PPA score gain, defined as the reduction in the whole-design PPA score relative to the parent configuration:
\begin{equation}
    g_t(r,b)=S(D_t)-S(D'_{t,r,b}).
\end{equation}
Using the regional embedding $z_r$ and branch $b\in\{\mathrm{RTL},\mathrm{SYN}\}$ as inputs, the GP directly models the PPA score gain as
\begin{equation}
    g_t(r,b)\mid f(z_r,b)\sim
    \mathcal{N}\!\left(f(z_r,b),\sigma_n^2\right),
    \label{eq:branch-surrogate}
\end{equation}
where $f$ is the latent GP gain function and $\sigma_n^2$ denotes the observation-noise variance.
To relate optimization feedback across structurally similar regions to guide branch selection, we use a Mat\'ern-$5/2$ covariance $k_{5/2}$ over their graph embeddings:
\begin{equation}
    \begin{aligned}
        k((z,b),(z',b'))&=k_{5/2}(\delta)\,\mathbf{1}[b=b'],\\
        \delta&=\frac{\lVert z-z'\rVert_2}{\ell}.
    \end{aligned}
    \label{eq:branch-kernel}
\end{equation}
Here, $\ell>0$ is the kernel length scale.
The GP thus predicts the PPA score gain from optimizing a region through a given branch, providing numerical guidance based on PPA feedback from optimizing regions with similar structures.

BO uses the GP posterior to balance exploitation and exploration when proposing region--branch pairs.
Let $\mu(z_r,b)$ and $v(z_r,b)$ denote the posterior mean and variance of the latent gain $f(z_r,b)$.
For each feasible region--branch pair $(r,b)$, the upper confidence bound is
\begin{equation}
    a(r,b)=\mu(z_r,b)+\kappa\sqrt{v(z_r,b)},
    \label{eq:branch-acquisition}
\end{equation}
where $\kappa$ controls the exploration--exploitation trade-off.
For each candidate region, BO passes its proposed branch and any available acquisition scores to the LLM Router.

The LLM Router also receives the region's RTL fragment and summaries of its CDFG and AIG.
To assess the BO proposal, it examines the region's RTL operations, control and data dependencies, and Boolean structure.
It combines this analysis with the available BO guidance to make the final branch assignment for each selected region, choosing either RTL rewriting or synthesis algorithm optimization.
Before optimization feedback is available, the LLM Router bases this choice on circuit semantics and structure.

The selected region--branch pairs are passed to the subsequent co-optimization stage, where each region is optimized through its assigned branch.
After evaluating each candidate, we update the GP with its PPA gain $g_t(r,b)$, the region's pre-optimization embedding $z_r$, and branch $b$.
The updated gain predictions and uncertainty inform the acquisition values used in subsequent rounds.

The next section describes how the RTL and SYN workers modify the regional RTL source or synthesis algorithms, respectively, and how the resulting changes are integrated into the current design configuration.





\section{RTL and Synthesis Algorithm Co-Optimization}
\label{sec:optimization}
Following bottleneck analysis, self-improving LLM workers optimize the selected regions through their assigned RTL or SYN branches, as illustrated in~\Cref{fig:method-overview}~(c).
At round $t$, all regional optimizations start from the same parent configuration $D_t=(R_t,A_t)$.
We then greedily integrate optimized RTL code and synthesis algorithm changes for regions, as shown in~\Cref{fig:method-overview}~(d).
The resulting design configuration $D_{t+1}$ produces the netlist and PPA reports for the next round.

\begin{table*}[!t]
    \centering
    \caption{PPA results of the initial design configurations, RTLScout~\cite{COOPT-ARXIV2026-Arnold}, Dr.\ RTL~\cite{RTL-ARXIV2026-Fang}, Self-Evolved ABC~\cite{LS-DAC2026-Yu}, and our method. WNS and TNS are in ps, area in $\mu\mathrm{m}^{2}$, and power in $\mu$W. Bold indicates the best results among the optimization methods, determined using unrounded values. Larger WNS/TNS and smaller area/power are better. For each design, let $x_0$ be its initial value and $x$ the method's result. Improvement (\%) is calculated as $100(x-x_0)/|x_0|$ for WNS/TNS and $100(x_0-x)/x_0$ for area/power, then averaged over the 16 designs. }
    \label{tab:initial-ppa}
    \scriptsize
    \fontfamily{lmr}\selectfont
    \setlength{\tabcolsep}{1.3pt}
    \resizebox{\textwidth}{!}{%
    \begin{tabular}{@{}l*{4}{r}|*{4}{r}|*{4}{r}|*{4}{r}|*{4}{r}@{}}
        \toprule
        \multirow{2}{*}{Design} & \multicolumn{4}{c|}{Initial}
        & \multicolumn{4}{c|}{RTLScout~\cite{COOPT-ARXIV2026-Arnold}} & \multicolumn{4}{c|}{Dr.\ RTL~\cite{RTL-ARXIV2026-Fang}}
        & \multicolumn{4}{c|}{Self-Evolved ABC~\cite{LS-DAC2026-Yu}} & \multicolumn{4}{c}{\textbf{Ours}} \\
        \cmidrule(lr){2-5}\cmidrule(lr){6-9}\cmidrule(lr){10-13}\cmidrule(lr){14-17}\cmidrule(lr){18-21}
        & WNS $\uparrow$ & TNS $\uparrow$ & Area $\downarrow$ & Power $\downarrow$
        & WNS $\uparrow$ & TNS $\uparrow$ & Area $\downarrow$ & Power $\downarrow$
        & WNS $\uparrow$ & TNS $\uparrow$ & Area $\downarrow$ & Power $\downarrow$
        & WNS $\uparrow$ & TNS $\uparrow$ & Area $\downarrow$ & Power $\downarrow$
        & WNS $\uparrow$ & TNS $\uparrow$ & Area $\downarrow$ & Power $\downarrow$ \\
        \midrule
        \texttt{hsm} & $-2.0\mathrm{E}3$ & $-3.7\mathrm{E}5$ & $6.3\mathrm{E}2$ & $6.7\mathrm{E}2$ & $-2.0\mathrm{E}3$ & $-3.7\mathrm{E}5$ & \textbf{6.3E2} & $6.7\mathrm{E}2$ & $-1.5\mathrm{E}3$ & $-3.3\mathrm{E}5$ & $6.4\mathrm{E}2$ & $6.7\mathrm{E}2$ & $-3.7\mathrm{E}3$ & $-5.9\mathrm{E}5$ & $6.6\mathrm{E}2$ & $6.7\mathrm{E}2$ & \textbf{-7.0E2} & \textbf{-1.1E5} & $6.3\mathrm{E}2$ & \textbf{5.1E2} \\
        \texttt{DSP} & $-6.7\mathrm{E}2$ & $-1.3\mathrm{E}4$ & $3.7\mathrm{E}2$ & $3.0\mathrm{E}2$ & $-2.4\mathrm{E}2$ & $-2.8\mathrm{E}3$ & \textbf{3.5E2} & $3.1\mathrm{E}2$ & $3.1\mathrm{E}2$ & \textbf{0} & $3.6\mathrm{E}2$ & $3.0\mathrm{E}2$ & $-2.4\mathrm{E}1$ & $-1.2\mathrm{E}2$ & $4.5\mathrm{E}2$ & $3.1\mathrm{E}2$ & \textbf{3.3E2} & \textbf{0} & $3.6\mathrm{E}2$ & \textbf{2.5E2} \\
        \texttt{aes} & $-4.9\mathrm{E}2$ & $-3.2\mathrm{E}5$ & $2.8\mathrm{E}3$ & $5.5\mathrm{E}3$ & $-9.8\mathrm{E}1$ & $-6.1\mathrm{E}4$ & $3.0\mathrm{E}3$ & $4.0\mathrm{E}3$ & $-3.6\mathrm{E}2$ & $-2.5\mathrm{E}5$ & $2.7\mathrm{E}3$ & $5.3\mathrm{E}3$ & $-9.8\mathrm{E}2$ & $-2.9\mathrm{E}5$ & $2.7\mathrm{E}3$ & $5.2\mathrm{E}3$ & \textbf{1.1E2} & \textbf{0} & \textbf{2.5E3} & \textbf{3.7E3} \\
        \texttt{arm\_cpu1} & $-5.2\mathrm{E}3$ & $-4.5\mathrm{E}6$ & $2.1\mathrm{E}3$ & $1.0\mathrm{E}3$ & $-1.6\mathrm{E}3$ & $-4.1\mathrm{E}5$ & $2.1\mathrm{E}3$ & $1.0\mathrm{E}3$ & $-1.3\mathrm{E}3$ & $-2.2\mathrm{E}5$ & $2.1\mathrm{E}3$ & $1.0\mathrm{E}3$ & $-1.4\mathrm{E}3$ & $-7.0\mathrm{E}5$ & \textbf{2.0E3} & $1.0\mathrm{E}3$ & \textbf{-4.7E2} & \textbf{-4.1E4} & $2.0\mathrm{E}3$ & \textbf{8.9E2} \\
        \texttt{arm\_cpu2} & $-9.8\mathrm{E}2$ & $-9.3\mathrm{E}4$ & $8.1\mathrm{E}2$ & $7.8\mathrm{E}2$ & $-2.7\mathrm{E}2$ & $-1.1\mathrm{E}4$ & $8.0\mathrm{E}2$ & $7.8\mathrm{E}2$ & $-3.5\mathrm{E}2$ & $-3.9\mathrm{E}4$ & $8.1\mathrm{E}2$ & $7.8\mathrm{E}2$ & $-1.1\mathrm{E}3$ & $-1.9\mathrm{E}5$ & $8.2\mathrm{E}2$ & $7.8\mathrm{E}2$ & \textbf{1.1E2} & \textbf{0} & \textbf{8.0E2} & \textbf{6.6E2} \\
        \texttt{controller} & $-2.1\mathrm{E}2$ & $-3.4\mathrm{E}3$ & $1.5\mathrm{E}1$ & $6.7\mathrm{E}1$ & $-7.4\mathrm{E}1$ & $-1.3\mathrm{E}3$ & $1.5\mathrm{E}1$ & $7.0\mathrm{E}1$ & $-9.0\mathrm{E}1$ & $-1.3\mathrm{E}3$ & $1.4\mathrm{E}1$ & \textbf{6.4E1} & $-1.5\mathrm{E}2$ & $-2.8\mathrm{E}3$ & $1.5\mathrm{E}1$ & $6.8\mathrm{E}1$ & \textbf{-6.7E1} & \textbf{-8.9E2} & \textbf{1.4E1} & $7.0\mathrm{E}1$ \\
        \texttt{cpu\_fsm} & $-8.9\mathrm{E}3$ & $-2.0\mathrm{E}5$ & $2.3\mathrm{E}3$ & $2.2\mathrm{E}3$ & $1.7\mathrm{E}4$ & \textbf{0} & $2.3\mathrm{E}3$ & $2.2\mathrm{E}3$ & $-8.8\mathrm{E}3$ & $-2.0\mathrm{E}5$ & $2.3\mathrm{E}3$ & $2.2\mathrm{E}3$ & $-7.0\mathrm{E}3$ & $-1.2\mathrm{E}5$ & $2.3\mathrm{E}3$ & $2.2\mathrm{E}3$ & \textbf{1.8E4} & \textbf{0} & \textbf{2.1E3} & \textbf{6.6E2} \\
        \texttt{cpu\_pipe} & $-3.2\mathrm{E}2$ & $-1.8\mathrm{E}4$ & $4.6\mathrm{E}2$ & $5.8\mathrm{E}2$ & $-3.2\mathrm{E}2$ & $-1.8\mathrm{E}4$ & $4.6\mathrm{E}2$ & $5.8\mathrm{E}2$ & $-2.8\mathrm{E}2$ & $-1.5\mathrm{E}4$ & $4.6\mathrm{E}2$ & $5.8\mathrm{E}2$ & $-9.6\mathrm{E}2$ & $-1.9\mathrm{E}5$ & $4.8\mathrm{E}2$ & $5.7\mathrm{E}2$ & \textbf{-2.6E2} & \textbf{-7.9E3} & \textbf{4.5E2} & \textbf{5.6E2} \\
        \texttt{datapath} & $-4.8\mathrm{E}2$ & $-1.9\mathrm{E}5$ & $8.7\mathrm{E}2$ & $1.6\mathrm{E}3$ & $-4.8\mathrm{E}2$ & $-1.9\mathrm{E}5$ & $8.7\mathrm{E}2$ & $1.6\mathrm{E}3$ & $-4.8\mathrm{E}2$ & $-1.3\mathrm{E}5$ & $8.6\mathrm{E}2$ & $1.7\mathrm{E}3$ & $-5.4\mathrm{E}2$ & $-1.9\mathrm{E}5$ & \textbf{8.5E2} & $1.6\mathrm{E}3$ & \textbf{2.3E2} & \textbf{0} & $8.7\mathrm{E}2$ & \textbf{1.5E3} \\
        \texttt{i2c} & $-1.9\mathrm{E}2$ & $-4.9\mathrm{E}3$ & $9.0\mathrm{E}1$ & $4.3\mathrm{E}2$ & $-1.5\mathrm{E}2$ & $-4.3\mathrm{E}3$ & \textbf{8.9E1} & $4.3\mathrm{E}2$ & $-8.3\mathrm{E}1$ & $-2.2\mathrm{E}3$ & $9.0\mathrm{E}1$ & $4.3\mathrm{E}2$ & $-1.0\mathrm{E}2$ & $-3.3\mathrm{E}3$ & $9.1\mathrm{E}1$ & $4.4\mathrm{E}2$ & \textbf{-8.2E1} & \textbf{-1.6E3} & $9.1\mathrm{E}1$ & \textbf{4.2E2} \\
        \texttt{router} & $-8.5\mathrm{E}2$ & $-3.0\mathrm{E}5$ & $3.6\mathrm{E}2$ & $1.6\mathrm{E}3$ & $-8.5\mathrm{E}2$ & $-3.0\mathrm{E}5$ & $3.6\mathrm{E}2$ & $1.6\mathrm{E}3$ & $-8.5\mathrm{E}2$ & $-3.0\mathrm{E}5$ & $3.6\mathrm{E}2$ & $1.6\mathrm{E}3$ & $-8.3\mathrm{E}2$ & $-1.3\mathrm{E}5$ & $3.7\mathrm{E}2$ & $1.6\mathrm{E}3$ & \textbf{-1.2E2} & \textbf{-9.3E3} & \textbf{3.6E2} & \textbf{1.6E3} \\
        \texttt{simple\_spi} & $-2.3\mathrm{E}2$ & $-6.8\mathrm{E}3$ & $7.9\mathrm{E}1$ & $5.7\mathrm{E}2$ & $-1.5\mathrm{E}2$ & $-5.5\mathrm{E}3$ & \textbf{7.7E1} & $5.7\mathrm{E}2$ & $-7.3\mathrm{E}1$ & $-1.9\mathrm{E}3$ & $7.8\mathrm{E}1$ & $5.7\mathrm{E}2$ & $-1.1\mathrm{E}2$ & $-3.9\mathrm{E}3$ & $8.1\mathrm{E}1$ & $5.7\mathrm{E}2$ & \textbf{-6.3E1} & \textbf{-1.7E3} & $7.8\mathrm{E}1$ & \textbf{5.7E2} \\
        \texttt{ticket\_machine} & $-1.4\mathrm{E}2$ & $-9.2\mathrm{E}2$ & $4.4\mathrm{E}0$ & $1.1\mathrm{E}2$ & $-7.0\mathrm{E}1$ & $-4.9\mathrm{E}2$ & $5.1\mathrm{E}0$ & $1.2\mathrm{E}2$ & $-7.6\mathrm{E}1$ & $-5.8\mathrm{E}2$ & $5.2\mathrm{E}0$ & $1.2\mathrm{E}2$ & $-9.7\mathrm{E}1$ & $-6.5\mathrm{E}2$ & $4.7\mathrm{E}0$ & $1.1\mathrm{E}2$ & \textbf{-4.4E1} & \textbf{-2.9E2} & \textbf{4.5E0} & \textbf{1.1E2} \\
        \texttt{fpadd} & $-1.4\mathrm{E}3$ & $-4.0\mathrm{E}4$ & $2.1\mathrm{E}2$ & $1.3\mathrm{E}2$ & $-7.3\mathrm{E}2$ & $-2.1\mathrm{E}4$ & $2.4\mathrm{E}2$ & $1.6\mathrm{E}2$ & $-3.2\mathrm{E}1$ & $-7.4\mathrm{E}2$ & \textbf{2.1E2} & $1.3\mathrm{E}2$ & $-3.6\mathrm{E}2$ & $-1.0\mathrm{E}4$ & $2.2\mathrm{E}2$ & $1.2\mathrm{E}2$ & \textbf{6.4E1} & \textbf{0} & $2.2\mathrm{E}2$ & \textbf{1.2E2} \\
        \texttt{fpmul} & $-9.8\mathrm{E}2$ & $-1.4\mathrm{E}4$ & $1.7\mathrm{E}2$ & $1.5\mathrm{E}2$ & $-7.7\mathrm{E}2$ & $-1.1\mathrm{E}4$ & \textbf{1.5E2} & \textbf{1.4E2} & $-3.7\mathrm{E}2$ & $-5.4\mathrm{E}3$ & $1.7\mathrm{E}2$ & $1.5\mathrm{E}2$ & $-8.1\mathrm{E}2$ & $-1.2\mathrm{E}4$ & $1.8\mathrm{E}2$ & $1.5\mathrm{E}2$ & \textbf{-5.0E1} & \textbf{-5.3E2} & $1.6\mathrm{E}2$ & $1.4\mathrm{E}2$ \\
        \texttt{pdpu} & $-3.3\mathrm{E}3$ & $-4.9\mathrm{E}4$ & $1.1\mathrm{E}3$ & $3.7\mathrm{E}2$ & $-1.0\mathrm{E}3$ & $-1.4\mathrm{E}4$ & $1.2\mathrm{E}3$ & $4.0\mathrm{E}2$ & $-2.7\mathrm{E}2$ & $-2.2\mathrm{E}3$ & $1.2\mathrm{E}3$ & $3.9\mathrm{E}2$ & $-2.8\mathrm{E}3$ & $-4.2\mathrm{E}4$ & \textbf{1.1E3} & $3.7\mathrm{E}2$ & \textbf{3.4E1} & \textbf{0} & $1.2\mathrm{E}3$ & \textbf{2.9E2} \\
        \midrule
        Average & $-1.6\mathrm{E}3$ & $-3.8\mathrm{E}5$ & $7.7\mathrm{E}2$ & $1.0\mathrm{E}3$ & $5.4\mathrm{E}2$ & $-8.8\mathrm{E}4$ & $7.9\mathrm{E}2$ & $9.2\mathrm{E}2$ & $-9.1\mathrm{E}2$ & $-9.4\mathrm{E}4$ & $7.7\mathrm{E}2$ & $1.0\mathrm{E}3$ & $-1.3\mathrm{E}3$ & $-1.6\mathrm{E}5$ & $7.7\mathrm{E}2$ & $9.8\mathrm{E}2$ & \textbf{1.0E3} & \textbf{-1.1E4} & \textbf{7.4E2} & \textbf{7.5E2} \\
        Improvement (\%) & --- & --- & --- & --- & $55.47$ & $44.96$ & $-2.43$ & $-1.22$ & $51.52$ & $50.97$ & $-1.52$ & $-1.01$ & $2.30$ & $-39.61$ & $-3.30$ & $0.09$ & \textbf{103.27} & \textbf{87.63} & \textbf{1.23} & \textbf{13.26} \\
        \bottomrule
    \end{tabular}%
    }
\end{table*}

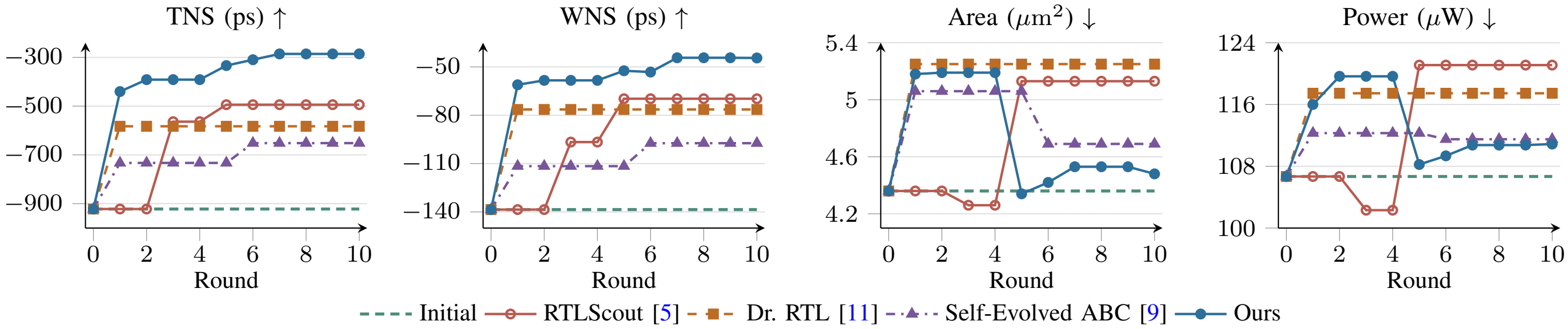
\begin{figure*}[!t]
    \centering
\newcommand{\plotTicketMachineMetric}[2]{%
    \addplot[plotTeal, densely dashed, line width=1pt] coordinates {(0,#2) (10,#2)};
    \addplot[plotRed, thick, mark=o, mark size=1.4pt]
        table[col sep=comma,x=round,y=rtlscout_#1] {fig/ticket-machine-trajectories.csv};
    \addplot[plotOrange, thick, dashed, mark=square*, mark size=1.4pt,
        mark options={solid}]
        table[col sep=comma,x=round,y=drrtl_#1] {fig/ticket-machine-trajectories.csv};
    \addplot[plotPurple, thick, dashdotted, mark=triangle*, mark size=1.6pt,
        mark options={solid}]
        table[col sep=comma,x=round,y=abc_#1] {fig/ticket-machine-trajectories.csv};
    \addplot[plotBlue, thick, mark=*, mark size=1.4pt]
        table[col sep=comma,x=round,y=ours_#1] {fig/ticket-machine-trajectories.csv};
}

\begin{tikzpicture}
    \begin{groupplot}[
        group style={group size=4 by 1, horizontal sep=0.074\textwidth},
        scale only axis,
        width=0.18\textwidth,
        height=2.15cm,
        axis lines=left,
        xmin=-0.3, xmax=10.3,
        xtick={0,2,4,6,8,10},
        xlabel={Round},
        x label style={at={(axis description cs:0.5,0)}, anchor=north,
            yshift=-0.39cm, font=\footnotesize},
        tick align=outside,
        tick label style={font=\footnotesize},
        xticklabel style={inner ysep=1.5pt},
        title style={font=\small, yshift=-5pt},
        scaled y ticks=false,
        ymajorgrids,
        grid style={gray!30, line width=0.3pt},
    ]
        \nextgroupplot[title={TNS (ps) $\uparrow$}, ymin=-1000, ymax=-240,
            ytick={-900,-700,-500,-300}]
        \plotTicketMachineMetric{tns}{-921.98}

        \nextgroupplot[title={WNS (ps) $\uparrow$}, ymin=-150, ymax=-35,
            ytick={-140,-110,-80,-50}]
        \plotTicketMachineMetric{wns}{-138.54}

        \nextgroupplot[title={Area ($\mu\mathrm{m}^{2}$) $\downarrow$},
            ymin=4.1, ymax=5.4, ytick={4.2,4.6,5.0,5.4},
            legend columns=5,
            legend style={at={(0,0)}, xshift=-0.037\textwidth,
                yshift=-0.73cm, anchor=north, draw=none, font=\footnotesize},
            legend entries={
                Initial,
                {RTLScout~\cite{COOPT-ARXIV2026-Arnold}},
                {Dr.\ RTL~\cite{RTL-ARXIV2026-Fang}},
                {Self-Evolved ABC~\cite{LS-DAC2026-Yu}},
                Ours
            }]
        \plotTicketMachineMetric{area}{4.36}

        \nextgroupplot[title={Power ($\mu$W) $\downarrow$},
            ymin=100, ymax=124, ytick={100,108,116,124}]
        \plotTicketMachineMetric{power}{106.6761106}
    \end{groupplot}
\end{tikzpicture}\par
    \vspace{-5pt}
    \caption{PPA trajectories for \texttt{ticket\_machine} at rounds 0--10. Our method achieves best values earlier than baselines.}
    \label{fig:ticket-machine-trajectories}
\end{figure*}

\subsection{RTL and Synthesis Algorithm Optimization with Self-Improving LLM Workers}
\label{sec:interventions}

Each worker uses its branch-specific optimization knowledge to optimize the regions assigned to it by the LLM Router.
Let $\mathcal{K}_t^b$ denote the optimization knowledge for branch $b$ at round $t$.
The RTL worker rewrites only the RTL source code implementing the selected region.
The SYN worker holds $R_t$ fixed and applies its modified synthesis algorithm only to the AIG subgraph of the selected region.
For each selected region--branch pair $(r,b)$, the worker uses this knowledge to generate a candidate configuration from the common parent $D_t$:
\begin{equation}
    D'_{t,r,b}=\operatorname{Worker}_b\!\left(D_t,r,\mathcal{K}_t^b\right).
    \label{eq:worker-configuration}
\end{equation}
Each candidate applies only the selected region's changes and undergoes functional equivalence checking before synthesis and whole-design PPA evaluation.
Its score gain $g_t(r,b)=S(D_t)-S(D'_{t,r,b})$ measures the effect of the regional optimization relative to the parent configuration.

As optimization proceeds, obtaining further score gains from previously optimized regions can become harder, motivating refinement of the workers' optimization knowledge.
Inspired by TextGrad~\cite{LLM-ARXIV2024-Yuksekgonul}, we treat $\mathcal{K}_t^b$ as a textual variable.
The reviewer uses the parent configuration, the regional transformations, and their PPA gains to generate \textbf{textual refinements} of the current knowledge.
\begin{equation}
    \mathcal{K}_{t+1}^b=\operatorname{Reviewer}\!\left(\mathcal{K}_t^b,D_t,\{(D'_{t,r,b},g_t(r,b))\}_r\right).
    \label{eq:knowledge-update}
\end{equation}
Here, $r$ ranges over the evaluated regions assigned to branch $b$ at round $t$.
The resulting natural-language feedback consolidates successful optimization strategies and suppresses ineffective ones based on their observed PPA outcomes.
The workers use the updated knowledge $\mathcal{K}_{t+1}^b$ in the next round.



\subsection{Design Configuration Update via Regional Integration}
\label{sec:integration}

As shown in~\Cref{fig:method-overview}~(d), after PPA evaluation, we select the $M$ candidate configurations $\{D'_{t,r_i,b_i}\}_{i=1}^M$ with positive PPA score gains for integration.
We next integrate the regional changes to form the next design configuration $D_{t+1}$.
However, combining regional changes can alter their effects on the synthesized design, so a change that improves the score on its own may lose that benefit after integration, given the NP-hard nature of synthesis~\cite{SYN-TCAD93-Li}.
In our setting, exhaustive integration can involve up to $2^M$ candidate combinations, with each feasible combination requiring whole-design synthesis and PPA evaluation.
To limit this evaluation cost, our \textbf{greedy} strategy considers candidates in descending order of their individual PPA gains and keeps each change only if it reduces the current whole-design PPA score.


For a candidate produced by optimizing region $r_i$ through branch $b_i$, the gain $g_t(r_i,b_i)=S(D_t)-S(D'_{t,r_i,b_i})$ determines the integration order:
\begin{equation}
    g_t(r_1,b_1)\geq g_t(r_2,b_2)\geq\cdots\geq g_t(r_M,b_M)>0.
    \label{eq:integration-order}
\end{equation}

We initialize the current configuration as $D_{\mathrm{cur}}=D_t$. At step $i$, $\operatorname{Integrate}(D_{\mathrm{cur}},r_i)$ applies the stored changes from $D'_{t,r_i,b_i}$ to form a trial configuration $D_{\mathrm{trial}}$. We synthesize the trial configuration to evaluate its PPA score.
The update is
\begin{equation}
    D_{\mathrm{cur}}\leftarrow
    \begin{cases}
        D_{\mathrm{trial}}, & \text{if } S(D_{\mathrm{trial}})<S(D_{\mathrm{cur}}),\\
        D_{\mathrm{cur}}, & \text{otherwise}.
    \end{cases}
    \label{eq:integration-update}
\end{equation}
After all candidates have been considered, we set $D_{t+1}=D_{\mathrm{cur}}$.
Its netlist and PPA reports support the next round of bottleneck analysis and co-optimization.
The next section presents the experimental evaluation of our co-optimization framework.

\section{Experimental Results}

\subsection{Experimental Setup}
We collect 16 RTL designs listed in~\Cref{tab:initial-ppa}, which contain an average of 1000 lines of RTL code and cover a wide range of applications, including CPU cores, DSPs, control circuits, etc.
We use ABC~\cite{LS-ABC-Brayton} to synthesize all designs with the ASAP7 PDK~\cite{PDK-ICCAD2017-Xu}.
We then evaluate the PPA of the resulting netlists using a commercial EDA tool.
For each design, we shorten the target clock period $T_{\mathrm{clk}}$ to obtain an initial WNS of approximately $-0.4T_{\mathrm{clk}}$.


In terms of the optimization objective, we consider four metrics: WNS, TNS, area, and power.
The score weights in~\Cref{eq:score} are fixed at
$(\omega_W,\omega_T,\omega_A,\omega_P)=(0.4,0.3,0.15,0.15)$ for WNS, TNS, area, and power, respectively.
We compare our framework with RTL-only optimization methods including Dr.\ RTL~\cite{RTL-ARXIV2026-Fang} and RTLScout~\cite{COOPT-ARXIV2026-Arnold}, and synthesis-only optimization methods including Self-Evolved ABC~\cite{LS-DAC2026-Yu}.
We run 10 optimization rounds and use DeepSeek-V4.1-Flash-1017 as the default backbone LLM for all methods.


\begin{figure*}[!t]
    \centering
    \includegraphics[width=\textwidth]{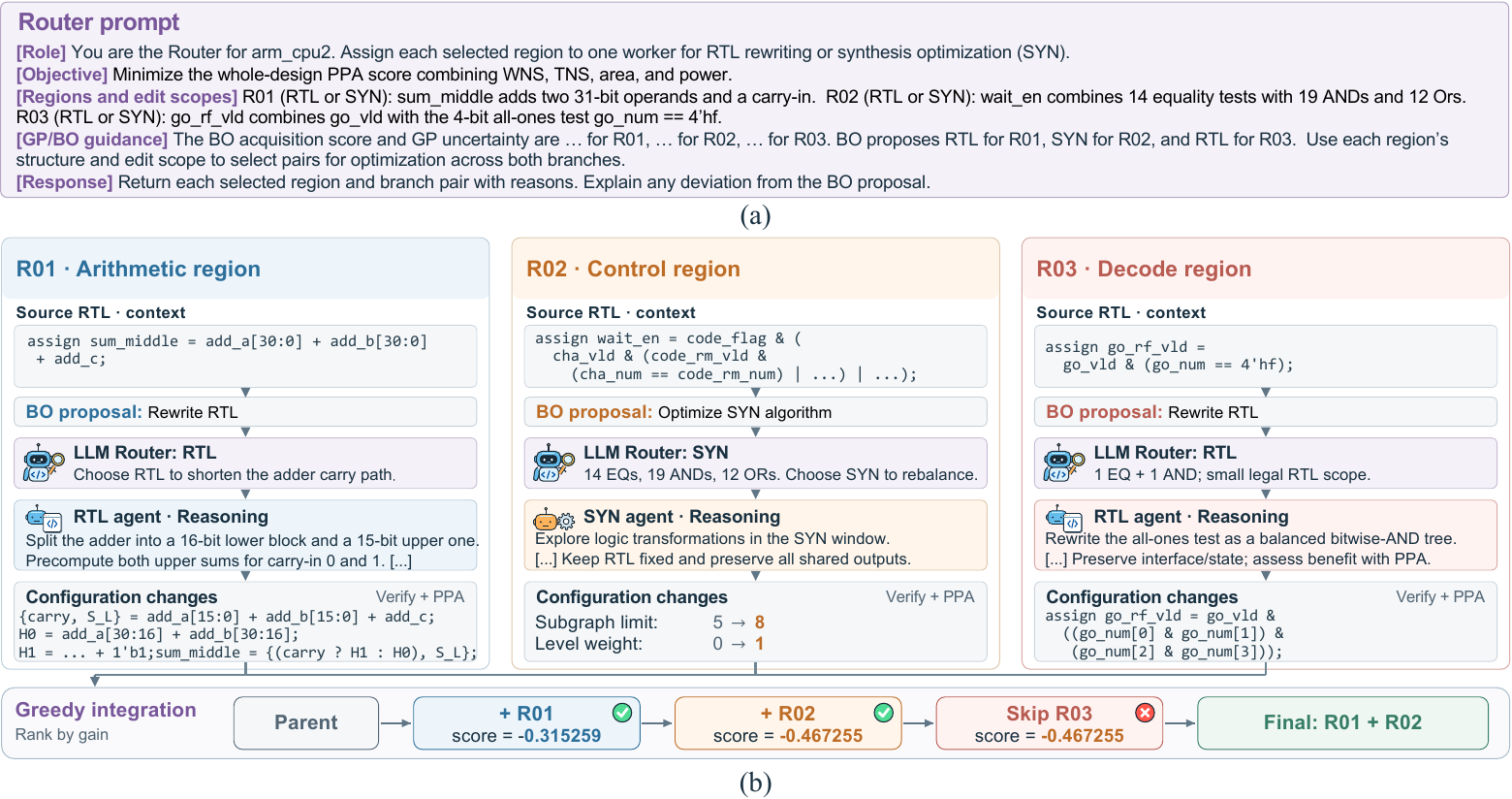}
    \caption{One round of co-optimization of 3 regions on \texttt{arm\_cpu2}: (a) LLM Router prompt; (b) illustration of region-branch assignments, co-optimization, and greedy integration procedures of regions R01, R02 and R03.}
    \label{fig:case-study}
    \vspace{-10pt}
\end{figure*}

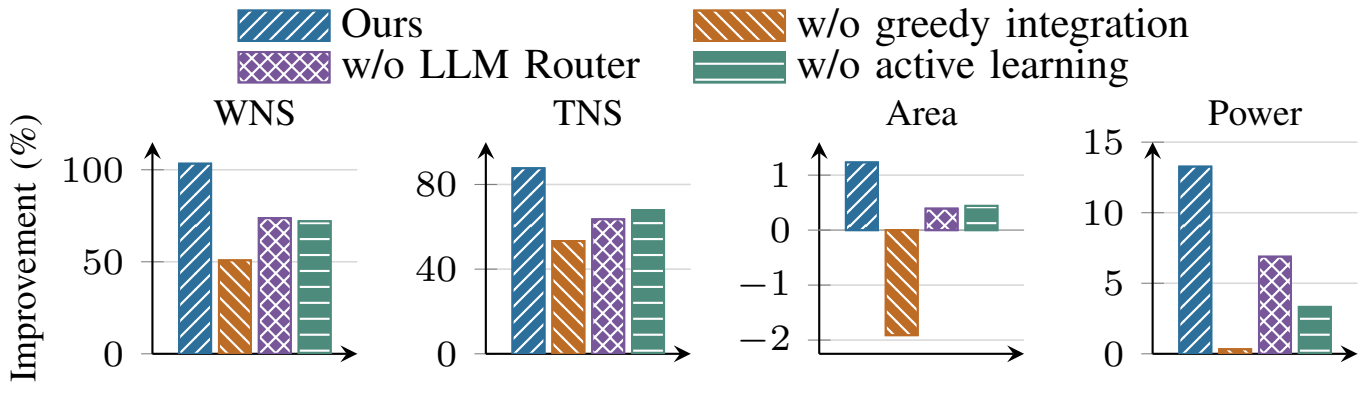
\begin{figure}[!t]
    \centering
\begin{tikzpicture}
    \newcommand{\plotAblationMetric}[4]{%
        \addplot[fill=plotBlue, draw=plotBlue,
            postaction={pattern=north east lines, pattern color=white}]
            coordinates {(0,#1)};
        \addplot[fill=plotOrange, draw=plotOrange,
            postaction={pattern=north west lines, pattern color=white}]
            coordinates {(0,#2)};
        \addplot[fill=plotPurple, draw=plotPurple,
            postaction={pattern=crosshatch, pattern color=white}]
            coordinates {(0,#3)};
        \addplot[fill=plotTeal, draw=plotTeal,
            postaction={pattern=horizontal lines, pattern color=white}]
            coordinates {(0,#4)};
    }
    \begin{groupplot}[
        group style={group size=4 by 1, horizontal sep=0.096\columnwidth},
        scale only axis,
        width=0.153\columnwidth,
        height=1.4cm,
        axis lines=left,
        ybar=1.5pt,
        area legend,
        /tikz/bar width=6pt,
        xmin=-0.5, xmax=0.5,
        xtick=\empty,
        ymin=0,
        scaled y ticks=false,
        ymajorgrids,
        grid style={gray!30, line width=0.3pt},
        tick label style={font=\scriptsize},
        yticklabel style={inner ysep=0pt},
        every axis title shift=3pt,
        title style={font=\scriptsize, inner ysep=0pt},
    ]
        \nextgroupplot[title={WNS}, ymax=115, ytick={0,50,100},
            ylabel={Improvement (\%)},
            ylabel style={font=\scriptsize, inner sep=0pt}]
        \plotAblationMetric{103.27}{50.75}{73.69}{72.00}

        \nextgroupplot[title={TNS}, ymax=100, ytick={0,40,80}]
        \plotAblationMetric{87.63}{53.26}{63.52}{67.90}

        \nextgroupplot[title={Area}, ymin=-2.25, ymax=1.6, ytick={-2,-1,1},
            extra y ticks={0},
            extra y tick style={grid=major, major grid style={gray!70}},
            legend columns=2,
            legend cell align=left,
            legend style={at={(0,1)}, xshift=-0.075\columnwidth,
                yshift=10pt, anchor=south, draw=none, font=\footnotesize,
                inner ysep=0pt, nodes={inner ysep=0.5pt},
                /tikz/every even column/.append style={column sep=8pt}},
            legend entries={Ours,w/o greedy integration,w/o LLM Router,w/o active learning}]
        \plotAblationMetric{1.23}{-1.91}{0.39}{0.44}

        \nextgroupplot[title={Power}, ymax=15, ytick={0,5,10,15}]
        \plotAblationMetric{13.26}{0.32}{6.88}{3.32}
    \end{groupplot}
\end{tikzpicture}
    \vspace{-10pt }
    \caption{Average improvements (\%) in WNS, TNS, area, and power relative to the initial design configurations.}
    \label{fig:ablation-average}
\end{figure}

\subsection{Experimental Results}
\Cref{tab:initial-ppa} compares the post-synthesis PPA results across all 16 designs.
Higher WNS and TNS indicate better timing, whereas lower power and area are preferable.
For each design, we compute the percentage improvement over the initial configuration as $100(x-x_0)/|x_0|$ for WNS and TNS, and $100(x_0-x)/x_0$ for power and area, where $x_0$ and $x$ denote the initial and optimized metric values, respectively.
Positive values indicate improvement, and we average these percentages over all 16 designs.

Across the 16 designs in~\Cref{tab:initial-ppa}, our framework achieves the highest WNS on every design and matches or exceeds the best baseline TNS on each one.
Relative to the initial configurations, our framework achieves average improvements of 103.27\% in WNS and 87.63\% in TNS.
These improvements exceed those of the strongest baseline for each timing metric by 47.80 percentage points for WNS (RTLScout~\cite{COOPT-ARXIV2026-Arnold}) and 36.66 percentage points for TNS (Dr.\ RTL~\cite{RTL-ARXIV2026-Fang}).

Across the 16 designs, our framework achieves the largest average improvements in all four metrics, although individual designs exhibit area and power tradeoffs.
On \texttt{DSP}, for example, RTLScout uses less area than our framework, while our framework consumes less power.
On \texttt{cpu\_fsm}, both RTLScout~\cite{COOPT-ARXIV2026-Arnold} and our framework achieve timing closure, but our implementation consumes 0.66~mW and occupies approximately 2,100~$\mu\mathrm{m}^{2}$, compared with 2.2~mW and 2,300~$\mu\mathrm{m}^{2}$ for RTLScout.
Relative to the initial configurations, our framework reduces area by 1.23\% and power by 13.26\% on average. RTLScout~\cite{COOPT-ARXIV2026-Arnold}, Dr.\ RTL~\cite{RTL-ARXIV2026-Fang}, and Self-Evolved ABC~\cite{LS-DAC2026-Yu} increase area by 2.43\%, 1.52\%, and 3.30\% on average, respectively.
These results support the effectiveness of performing co-optimization with RTL rewriting and synthesis algorithm optimization to improve overall PPA.


To illustrate the efficiency of our framework, we compare the optimization trajectories of \texttt{ticket\_machine} in~\Cref{fig:ticket-machine-trajectories}.
The timing advantage appears in the first round. 
Our framework reaches a WNS of $-61.10$~ps and a TNS of $-440.31$~ps, already surpassing the best final baseline values of $-69.76$~ps and $-494.14$~ps, both achieved by RTLScout~\cite{COOPT-ARXIV2026-Arnold}.
By round~5, our design occupies 4.34~$\mu\mathrm{m}^{2}$ and consumes 108.23~$\mu$W, both below the best final baseline values achieved by Self-Evolved ABC~\cite{LS-DAC2026-Yu}.
From round~5 onward, our framework outperforms the final configurations of all three baselines on all four metrics.

\subsection{Case Study}
\Cref{fig:case-study} shows how bottleneck analysis guides the choice between RTL rewriting and synthesis algorithm optimization for three regions of \texttt{arm\_cpu2} and how the optimized regions are integrated in a single optimization round.
BO proposes an optimization branch for each region.
The LLM Router reviews these proposals using the prompt shown in~\Cref{fig:case-study}~(a) and assigns each region to an RTL or SYN worker.
In this round, the RTL workers rewrite the arithmetic region R01 and the decode region R03, and the SYN worker optimizes the synthesis algorithm for the control region R02.
Regions R01 and R02 are the two regions shown in~\Cref{fig:optimization-overview}.
After RTL rewriting, the AIG depth of R01 decreases from 65 to 45 levels. Synthesis optimization reduces the AIG depth of R02 from 13 to 11 levels.
Greedy integration ranks the candidates by their individual gains.
In~\Cref{fig:case-study}~(b), the current configuration has a whole-design PPA score of $-0.315259$ after R01 is accepted and $-0.467255$ after R02 is added.
Expanding the word-level comparison in R03 into bit-level logic leaves the synthesized implementation unchanged, so R03 is skipped and the score remains $-0.467255$.
Overall, our framework addresses regional bottlenecks through RTL and synthesis co-optimization with bottleneck analysis and converts regional improvements into whole-design PPA gains through greedy integration.

\subsection{Ablation Study}
To assess each component's contribution to PPA, we remove greedy integration, the LLM Router, and active learning from our framework one at a time.
Among the four configurations in~\Cref{fig:ablation-average}, the ablation experiment without greedy integration yields the smallest average gains in WNS, TNS, and power.
Removing greedy integration also shifts the average area change from a 1.23\% reduction to a 1.91\% increase. This result suggests that regional changes that improve PPA individually can lose their benefit when integrated.
Removing either the LLM Router or active learning also reduces gains across all four metrics. For example, the average WNS improvement falls from 103.27\% with the full framework to 73.69\% without the Router and 72.00\% without active learning. These results suggest that a region's semantics help the Router identify which branch is better suited to its bottleneck, while PPA feedback from earlier rounds with structural embeddings helps it recognize promising region--branch pairs.
Together, these components enable the co-optimization framework assign regions with RTL rewriting or synthesis optimization and refine subsequent assignments as it learns which choices improve PPA on the design.

\section{Conclusion}
\label{sec:conclusion}
In this paper, we present an LLM-assisted framework for co-optimization of RTL and synthesis algorithms to improve post-synthesis PPA.
Bottleneck analysis assesses whether RTL rewriting or synthesis algorithm optimization is more likely to improve regions in design that lead to sub-optimal PPA.
This analysis combines LLM reasoning over regional CDFG and AIG structures with GP-guided Bayesian optimization informed by accumulated PPA feedback.
Guided by the analysis, self-improving RTL and SYN workers address the identified bottlenecks through RTL rewriting and synthesis algorithm optimization, respectively.
Their regional changes are then greedily integrated into the current design configuration under whole-design PPA evaluation.
Across 16 RTL designs, our framework achieves the best WNS, TNS, area, and power compared with RTL-only and synthesis-only methods.
By optimizing both RTL and synthesis algorithms with bottleneck analysis, this work takes a step toward the broader vision of co-optimizing hardware designs and EDA algorithms.


\clearpage
{
    \bibliographystyle{IEEEtran}
    \bibliography{ref/Top-sim, ref/Synthesis, ref/Gen, ref/OPT, ref/Additional, ref/Paper2}

@article{LLM-ARXIV2024-Yuksekgonul,
  title={{TextGrad}: Automatic ``Differentiation'' via Text},
  author={Yuksekgonul, Mert and Bianchi, Federico and Boen, Joseph and Liu, Sheng and Huang, Zhi and Guestrin, Carlos and Zou, James},
  journal={arXiv preprint arXiv:2406.07496},
  year={2024},
  doi={10.48550/arXiv.2406.07496},
  url={https://arxiv.org/abs/2406.07496}
}

@misc{RTL-ARXIV2026-Fang,
  title = {{Dr. RTL}: Autonomous Agentic {RTL} Optimization through Tool-Grounded Self-Improvement},
  author = {Fang, Wenji and Lu, Yao and Liu, Shang and Wang, Jing and Guo, Ziyan and He, Junxian and Tu, Fengbin and Xie, Zhiyao},
  year = {2026},
  eprint = {2604.14989},
  archivePrefix = {arXiv},
  primaryClass = {cs.AI},
  doi = {10.48550/arXiv.2604.14989},
}

@misc{COOPT-ARXIV2026-Arnold,
  title = {{RTLScout}: Joint Agentic Code and Synthesis Optimization for Efficient Digital Circuits},
  author = {Arnold, Felix and Amaudruz, Ryan and Tsaras, Dimitrios and Andri, Renzo and Cavigelli, Lukas},
  year = {2026},
  eprint = {2606.06530},
  archivePrefix = {arXiv},
  primaryClass = {cs.AR},
  doi = {10.48550/arXiv.2606.06530},
}

@inproceedings{LS-DAC2006-Mishchenko,
  title = {{DAG}-Aware {AIG} Rewriting: A Fresh Look at Combinational Logic Synthesis},
  author = {Mishchenko, Alan and Chatterjee, Satrajit and Brayton, Robert},
  booktitle = dac,
  year = {2006},
  pages = {532--535},
  doi = {10.1145/1146909.1147048}
}

@inproceedings{LS-CAV2010-Brayton,
  title = {{ABC}: An Academic Industrial-Strength Verification Tool},
  author = {Brayton, Robert and Mishchenko, Alan},
  booktitle = cav,
  year = {2010},
  pages = {24--40},
  publisher = {Springer},
  doi = {10.1007/978-3-642-14295-6_5}
}

@article{RTL-TCAD2024-Coward,
  title = {{ROVER}: {RTL} Optimization via Verified E-Graph Rewriting},
  author = {Coward, Samuel and Drane, Theo and Constantinides, George A.},
  journal = tcad,
  year = {2024},
  volume = {43},
  number = {12},
  pages = {4687--4700},
  doi = {10.1109/TCAD.2024.3410154},
}

@inproceedings{RTL-ICCAD2024-Yao,
  title = {{RTLRewriter}: Methodologies for Large Models aided {RTL} Code Optimization},
  author = {Yao, Xufeng and Wang, Yiwen and Li, Xing and Lian, Yingzhao and Chen, Ran and Chen, Lei and Yuan, Mingxuan and Xu, Hong and Yu, Bei},
  booktitle = iccad,
  year = {2024},
  articleno = {98},
  doi = {10.1145/3676536.3676775},
}

@inproceedings{LS-DAC2026-Yu,
  title = {Autonomous Evolution of {EDA} Tools: Multi-Agent Self-Evolved {ABC}},
  author = {Yu, Cunxi and Ren, Haoxing},
  booktitle = dac,
  year = {2026},
  publisher = {ACM},
  numpages = {7},
  doi = {10.1145/3770743.3804221},
}

@inproceedings{LS-ICCAD2007-Mishchenko,
  title = {Combinational and Sequential Mapping with Priority Cuts},
  author = {Mishchenko, Alan and Cho, Sungmin and Chatterjee, Satrajit and Brayton, Robert},
  booktitle = iccad,
  year = {2007},
  pages = {354--361},
  doi = {10.1109/ICCAD.2007.4397290},
}

@inproceedings{LS-DATE2019-Riener,
  title = {On-the-fly and {DAG}-aware: Rewriting Boolean Networks with Exact Synthesis},
  author = {Riener, Heinz and Haaswijk, Winston and Mishchenko, Alan and De Micheli, Giovanni and Soeken, Mathias},
  booktitle = date,
  year = {2019},
  pages = {1649--1654},
  doi = {10.23919/DATE.2019.8715185},
}

@inproceedings{RTL-ICML2026-Liu,
  title = {Beyond Tokens: Enhancing {RTL} Quality Estimation via Structural Graph Learning},
  author = {Liu, Yi and Zhang, Hongji and Wang, Yiwen and Tsaras, Dimitris and Chen, Lei and Yuan, Mingxuan and Xu, Qiang},
  booktitle = icml,
  year = {2026},
  volume = {306},
  series = {Proceedings of Machine Learning Research},
  pages = {79855--79876},
  publisher = {PMLR},
}

@inproceedings{RTL-AAAI2026-Ma,
  title = {{DynamicRTL}: {RTL} Representation Learning for Dynamic Circuit Behavior},
  author = {Ma, Ruiyang and Zhou, Yunhao and Wang, Yipeng and Liu, Yi and Shi, Zhengyuan and Zheng, Ziyang and Chen, Kexin and He, Zhiqiang and Yan, Lingwei and Chen, Gang and Xu, Qiang and Luo, Guojie},
  booktitle = aaai,
  year = {2026},
  volume = {40},
  number = {1},
  pages = {836--843},
  doi = {10.1609/aaai.v40i1.37051},
}

@inproceedings{RTL-DAC2023-Coward,
  title = {Automating Constraint-Aware Datapath Optimization using E-Graphs},
  author = {Coward, Samuel and Constantinides, George A. and Drane, Theo},
  booktitle = dac,
  year = {2023},
  pages = {1--6},
  doi = {10.1109/DAC56929.2023.10247797},
}

@inproceedings{RTL-NEURIPS2025-Wang,
  title = {{SymRTLO}: Enhancing {RTL} Code Optimization with {LLMs} and Neuron-Inspired Symbolic Reasoning},
  author = {Wang, Yiting and Ye, Wanghao and Guo, Ping and He, Yexiao and Wang, Ziyao and Tian, Bowei and He, Shwai and Sun, Guoheng and Shen, Zheyu and Chen, Sihan and Srivastava, Ankur and Zhang, Qingfu and Qu, Gang and Li, Ang},
  booktitle = neurips,
  year = {2025},
  volume = {38},
  pages = {50093--50118},
  publisher = {Curran Associates, Inc.},
  doi = {10.52202/085713-1670},
}

@inproceedings{LS-ASPDAC2022-Riener,
  title = {Boolean Rewriting Strikes Back: Reconvergence-Driven Windowing Meets Resynthesis},
  author = {Riener, Heinz and Lee, Siang-Yun and Mishchenko, Alan and De Micheli, Giovanni},
  booktitle = aspdac,
  year = {2022},
  pages = {395--402},
  doi = {10.1109/ASP-DAC52403.2022.9712526},
}

@misc{PD-ARXIV2026-Liu,
  title = {{GoalEvolve}: From Handcrafted Algorithm Priors to Goal-Driven Evolution of Physical Design Algorithms},
  author = {Liu, Haixu and Zhou, Lei and Ren, Yuhao and Wu, Yumao and Wang, Zhiang},
  year = {2026},
  eprint = {2608.16733},
  archivePrefix = {arXiv},
  doi = {10.48550/arXiv.2608.16733},
}

@inproceedings{EDA-DAC2026-Ghose,
  title = {Special Research Session: Progress and Roadmap Toward an Automated {EDA} {R\&D} Engineer},
  author = {Ghose, Amur and Kahng, Andrew B. and Pramanik, Bodhisatta},
  booktitle = dac,
  year = {2026},
  publisher = {ACM},
  numpages = {4},
  doi = {10.1145/3770743.3815396},
}

@misc{PD-ARXIV2026-Zhu,
  title = {Provably Good {Prim-Dijkstra} Revisited: New Theory and a Practical Algorithm for a Classical {VLSI} Routing Problem with {LLMs}},
  author = {Zhu, Keren},
  year = {2026},
  eprint = {2607.17005},
  archivePrefix = {arXiv},
  primaryClass = {cs.AR},
  doi = {10.48550/arXiv.2607.17005},
}

@article{ADDER-TCAD2019-Ma,
    title     = {Cross-layer Optimization for High Speed Adders: A Pareto Driven Machine Learning Approach},
    author    = {Ma, Yuzhe and Roy, Subhendu and Miao, Jin and Chen, Jiamin and Yu, Bei},
    journal   = tcad,
    year      = {2019}
}

@article{SYN-TCAD93-Li,
  title={On the circuit implementation problem},
  author={Li, W-N and Lim, Andrew and Agrawal, Prathima and Sahni, Sartaj},
  journal={IEEE transactions on computer-aided design of integrated circuits and systems},
  volume={12},
  number={8},
  pages={1147--1156},
  year={1993},
  publisher={IEEE}
}

@inproceedings{PDK-ICCAD2017-Xu,
  title={Standard cell library design and optimization methodology for ASAP7 PDK},
  author={Xu, Xiaoqing and Shah, Nishi and Evans, Andrew and Sinha, Saurabh and Cline, Brian and Yeric, Greg},
  booktitle=iccad,
  pages={999--1004},
  year={2017}
}

@inproceedings{RTL-ISCA2022-Xu,
  title={SNS's not a synthesizer: a deep-learning-based synthesis predictor},
  author={Xu, Ceyu and Kjellqvist, Chris and Wills, Lisa Wu},
  booktitle=isca,
  pages={847--859},
  year={2022}
}

@inproceedings{RTL-ICCAD2023-Fang,
  title={MasterRTL: A Pre-Synthesis PPA Estimation Framework for Any RTL Design},
  author={Fang, Wenji and Lu, Yao and Liu, Shang and Zhang, Qijun and Xu, Ceyu and Wills, Lisa Wu and Zhang, Hongce and Xie, Zhiyao},
  booktitle=iccad,
  pages={1--9},
  year={2023},
  organization={IEEE}
}

@inproceedings{LS-ABC-Brayton,
  title={ABC: An academic industrial-strength verification tool},
  author={Brayton, Robert and Mishchenko, Alan},
  booktitle={International Conference on Computer Aided Verification},
  pages={24--40},
  year={2010},
  organization={Springer}
}

@string{aspdac   = "Proc.~ASPDAC"}

@string{cav      = "Proc.~CAV"}

@string{dac      = "Proc.~DAC"}

@string{date     = "Proc.~DATE"}

@string{iccad    = "Proc.~ICCAD"}

@string{isca     = "Proc.~ISCA"}

@string{tcad     = "IEEE TCAD"}

@string{aaai     = "Proc.~AAAI"}

@string{icml     = "Proc.~ICML"}

@string{neurips  = "Proc.~NeurIPS"}

@string{arxiv    = "arXiv preprint"}
}

\end{document}